\documentclass
[superscriptaddress,secnumarabic,amssymb,amsmath,nobibnotes,aps,prd,showkeys,showpacs,nofootinbib,onecolumn]{revtex4}%
\usepackage{graphicx}
\usepackage{epsf}
\usepackage{bm}
\usepackage{amsmath}
\usepackage{amsfonts}
\usepackage{amssymb}
\usepackage{epstopdf}%
\providecommand{\U}[1]{\protect\rule{.1in}{.1in}}
\newcommand{\be}{\begin{equation}}
\newcommand{\ee}{\end{equation}}

\newcommand{\mincir}{\raise
-3.truept\hbox{\rlap{\hbox{$\sim$}}\raise4.truept\hbox{$<$}\ }}
\newcommand{\magcir}{\raise
-3.truept\hbox{\rlap{\hbox{$\sim$}}\raise4.truept\hbox{$>$}\ }}

\begin{document}
\title{Analytical solutions in $R+qR^{n}$ cosmology from singularity analysis}
\author{Andronikos Paliathanasis}
\email{anpaliat@phys.uoa.gr}
\affiliation{Instituto de Ciencias F\'{\i}sicas y Matem\'{a}ticas, Universidad Austral de
Chile, Valdivia, Chile}
\author{P.G.L. Leach}
\email{leach.peter@ucy.ac.cy}
\affiliation{Department of Mathematics and Institute of Systems Science, Research and
Postgraduate Support, Durban University of Technology, PO Box 1334, Durban
4000, Republic of South Africa}
\affiliation{School of Mathematics, Statistics and Computer Science, University of
KwaZulu-Natal, Private Bag X54001, Durban 4000, Republic of South Africa}
\affiliation{Department of Mathematics and Statistics, University of Cyprus, Lefkosia 1678, Cyprus}

\begin{abstract}
The integrability of higher-order theories of gravity is of importance in the
determining the properties of these models and so their viability as models of
reality. An important tool in the establishment of integrability is the
singularity analysis. We apply this analysis to the case of fourth-order
theory of gravity $f(R) = R + qR^{n}$ to establish those values of the free
parameters $q$ and $n$ for which integrability in this sense exists. As a
prelininary we examine the well-known case of $n = 4/3$.

\end{abstract}
\maketitle
\date{\today}
%

%TCIMACRO{\TeXButton{bigskip}{\bigskip}}%
%BeginExpansion
\bigskip
%EndExpansion

The generalization of the Einstein-Hilbert Action of General Relativity (GR),
which leads to the so-called modified theories of gravity, has been proposed
in order to explain some of the recent cosmological observations. The novelty
of the modified theories of gravity is that the dark energy, which is related
with the late-time acceleration of the universe, has its origins in the
additional dynamical quantities which follow from the gravitational action
\cite{capQ}.

A proposed generalization of the Einstein-Hilbert action is the so-called
$f\left(  R\right)  $-gravity in which the gravitational Action Integral, in a
four-dimensional Riemannian manifold $\mathit{M}$ with metric tensor
$g_{\mu\nu}\,$\ of Lorentzian signature, is given as follows \cite{Buda}%
\begin{equation}
S_{f\left(  R\right)  }=\int dx^{4}\sqrt{-g}f\left(  R\right)  +S_{m},
\label{ac.01}%
\end{equation}
where $R$ is the Ricci Scalar of the spacetime. The Action Integral,
(\ref{ac.01}), includes that of GR with or without cosmological constant when
$f\left(  R\right)  $ is a linear function, i.e., $f_{,RR}=0$. Furthermore the
Ricci Scalar admits second-order derivatives of the metric. However, when
$f\left(  R\right)  $ is a linear function, integration by parts of
(\ref{ac.01}) gives that the Lagrangian of the field equation admits only
first derivatives, that is, the gravitational theory is a second-order
gravity. However, the latter is not true for a nonlinear function, $f\left(
R\right)  $, where the theory is a fourth-order gravity\footnote{For reviews
in $f\left(  R\right)  $~theories of gravity see \cite{Sotiriou,odin1}.}.

It is well known that a fourth-order differential equation can be written as a
system of two second-order differential equations with the use of a Lagrange
Multiplier. The analogy in Classical Mechanics is the Legendre Transformation
between the Euler-Lagrange equations and Hamilton's equations. Furthermore
variation with respect to the metric tensor in (\ref{ac.01}) leads to the set
of equations,
\begin{equation}
G_{\mu\nu}^{f\left(  R\right)  }\equiv f_{,R}R_{\mu\nu}-\frac{1}{2}fg_{\mu\nu
}-\left(  \nabla_{\mu}\nabla_{\nu}-g_{\mu\nu}\nabla_{\sigma}\nabla^{\sigma
}\right)  f_{,R}=T_{\mu\nu}, \label{ac.02}%
\end{equation}
in which $T_{\mu\nu}$ is the energy-momentum tensor which is related with the
matter source, $S_{m}$ in (\ref{ac.01}), and $\nabla_{\mu}$ is the covariant
derivative associated with the Levi-Civita connection Riemannian manifold with
metric tensor $g_{\mu\nu}$ and $R_{\mu\nu}$ is the Ricci tensor field. Easily
we can see that for$~$nonlinear~function $f\left(  R\right)  $, the third term
of the left hand side of (\ref{ac.02}) provides the fourth-order derivatives.

However, if we assume the Ricci Scalar to be one of the dynamical variables of
the system, then only second-order derivatives exist in (\ref{ac.02}), while
the new equation which we have to consider is $R=R^{\mu\nu}R_{\mu\nu}$.
Another parametrization that has been proposed is to define the field
$\phi=f_{,R}$, where the field equations, (\ref{ac.02}), correspond to those
of Scalar-tensor theories and, specifically, to O'Hanlon gravity
\cite{Hanlon}. Furthermore the same results follow even if with a Lagrange
Multiplier $\lambda~$\cite{lan1}~and a new field $R~$or $\phi~$in the
gravitational action (\ref{ac.01}).

In the context of this work we are interested in the singularity analysis of
the field equations, (\ref{ac.02}), at the cosmological level. We select to
work on the reparameterization. in which the new dynamical variable that makes
the field equations a system of second-order differential equations is the
Ricci Scalar $R$. Furthermore we consider the metric $g_{\mu\nu}$ to be that
of a Friedmann-Lema\^{\i}tre-Robertson-Walker spacetime (FLRW) with zero
spatial curvature, which gives that the field equations (\ref{ac.02}) follow
from the variational principle of the Lagrangian function,%
\begin{equation}
L\left(  N,a,\dot{a},R,\dot{R}\right)  =\frac{1}{N}\left(  6af^{\prime}\dot
{a}^{2}+6a^{2}f^{\prime\prime}\dot{a}\dot{R}\right)  +N\left(  a^{3}\left(
f^{\prime}R-f\right)  +2\rho_{m0}\right)  ,\label{ac.03}%
\end{equation}
where we assumed the matter source to be that of dust fluid, $a\left(
t\right)  $ is the scale-factor and $N\left(  t\right)  $ is the
lapse-function; that is, the line element of the FLRW spacetime is%
\begin{equation}
ds^{2}=-N^{2}\left(  t\right)  dt^{2}+a^{2}\left(  t\right)  \left(
dx^{2}+dy^{2}+dz^{2}\right)  .\label{ac.04}%
\end{equation}

We select to work in the lapse in which $N\left(  t\right)  =1~$where the
comoving observer is $u^{\mu}=\delta_{t}^{\mu}$, $\left(  u^{\mu}u_{\mu
}=-1\right)  $. Therefore the field equations, (\ref{ac.02}),
are\footnote{$H=\dot{a}/a$ is the Hubble Function.}%
\begin{equation}
3f^{\prime}H^{2}+3Hf^{\prime\prime}\dot{R}-\left(  f^{\prime}R-f\right)
=\frac{\rho_{m_{0}}}{a^{3}} \label{fr.06}%
\end{equation}
and
\begin{equation}
2f^{\prime}\dot{H}+3f^{\prime}H^{2}=-2Hf^{\prime\prime}\dot{R}-\left(
f^{\prime\prime\prime}\dot{R}^{2}+f^{\prime\prime}\ddot{R}\right)
-\frac{f-Rf^{\prime}}{2}, \label{fr.07}%
\end{equation}
while the constraint equation $R=R^{\mu\nu}R_{\mu\nu}$ is
\begin{equation}
R=6\left(  \frac{\ddot{a}}{a}+\left(  \frac{\dot{a}}{a}\right)  ^{2}\right)  .
\label{fr.05}%
\end{equation}

A useful observation for equations (\ref{fr.06})-(\ref{fr.05}) is that they
form a Hamiltonian system, in which (\ref{fr.06}) is the Hamiltonian
function\footnote{The corresponding conservation law of the autonomous Noether
symmetry, $\partial_{t}$, of the Lagrangian function (\ref{ac.04}).}. That
means that the solution of the system, (\ref{fr.07}), (\ref{fr.05}), when it
is substituted into (\ref{fr.06}), provides a constraint equation for the free
parameters of the problem. \ Moreover, as has been discussed in \cite{palfr},
the field equation can be seen as the motion of particle in a two-dimensional
flat space in which the motion is driven by the potential, $V_{eff}\left(
a,R\right)  =a^{3}\left(  f^{\prime}R-R\right)  $. Hence for different
potentials, functions $f\left(  R\right)  $, the system has a different
evolution and that means that the scale factor, $a\left(  t\right)  $, evolves
differently and this corresponds to different cosmological models. Therefore
the functional form of $f\left(  R\right)  $ is essential.

Different models have been proposed in the literature, some of which can be
found in \cite{fr5,fr6,fr1,fr2} and references therein, while some models
which describe the accelerated expansion of the universe can be found in
\cite{odin2}. However, there are only few known forms for the $f\left(
R\right)  $ function in which the field equations form an integrable dynamical
system and consequently the solution can be written in a closed-form. Most of
these forms have been found with the method of group invariant transformations
\cite{palfr,christ,vakili,palcqg}, while the results of \cite{fer,ver1} on the
integrable scalar-tensor theories can be used in order to construct solutions
in $f\left(  R\right)  $-gravity.

In order to perform the singularity analysis of the field equations,
(\ref{fr.06})-(\ref{fr.05}), we consider that the function, $f\left(
R\right)  $, has the functional form \footnote{For the cosmological
implication of this model see \cite{amev}.} \cite{fr6}%
\begin{equation}
f\left(  R\right)  =R+qR^{n}, \label{ac.05}%
\end{equation}
where for $n=2$ the well-known Starobinsky model follows \cite{star}.
Furthermore we assume that $q\neq0$ and $n\neq0,1$. Otherwise the Action
Integral (\ref{ac.01}) is that of GR. Hence we determine constraints on the
power, $n$, in which the field equations, (\ref{fr.07})-(\ref{fr.05}), pass
the singularity test and the dynamical system is integrable in that the
solution can be written always as a Laurent expansion. Specifically for
\ (\ref{ac.05}) the Euler-Lagrange equations of (\ref{ac.03}) are equation
(\ref{fr.05}) and%
\begin{equation}
\ddot{R}+\frac{n-2}{R}\dot{R}^{2}+\frac{2}{a}\dot{a}\dot{R}-\frac{\left(
R+qnR^{n}\right)  }{qn\left(  n-1\right)  }R^{1-n}a^{-2}\dot{a}^{2}%
-\frac{\left(  qR^{n-1}\left(  n-3\right)  -2\right)  }{qn\left(  n-1\right)
}R^{1-n}=0, \label{ac.05a}%
\end{equation}
while the constraint equation (\ref{fr.06}) becomes
\begin{equation}
3\left(  1+nqR^{n-1}\right)  a\dot{a}^{2}+3n\left(  n-1\right)  a^{2}%
R^{n-2}\dot{a}\dot{R}-q\left(  n-1\right)  a^{3}R^{n}=\rho_{m0} \label{ac.05b}%
\end{equation}

The application of singularity analysis of differential equations in
gravitational studies has been used for the study of the integrability of the
field equations in GR for the Bianchi IX spacetime, the Mixmaster universe
\cite{Cotsakis,Con1,Demaret}, the singularity analysis of the Mixmaster
universe in a fourth-order gravity can be found in \cite{Cotsakis} while some
other applications in \cite{pb}.

For brevity we omit the basic properties of the singularity analysis which can
be found for instance in \cite{buntis,Feix97a,Andriopoulos06a} and references
therein, in the following the ARS algorithm is applied \cite{buntis}. Before
we start the analysis for the model (\ref{ac.05}) with general $n$, we
consider the well-known integrable case $n=\frac{4}{3}$, which has been found
with the application of Killing tensors in the minisuperspace approach
\cite{palcqg}.

We start by searching for the dominant behaviour, $\left(  a,R\right)
\rightarrow\left(  a_{0}\tau^{p},R_{0}\tau^{q}\right)  $, in the system
(\ref{fr.07})-(\ref{fr.05}) for $n=\frac{4}{3}~$,where $\tau=\left(
t-t_{0}\right)  $. From (\ref{fr.05}), as it is expected, we find that
$R_{0}=6\left(  2p^{2}-p\right)  $, while from the second one we find from the
dominant terms that%
\begin{equation}
p_{1}=0~,~p_{2}=\frac{1}{2}~,~p_{3}=\frac{5}{6}~,~p_{4}=\frac{8}{9}
\label{ac.06}%
\end{equation}
for arbitrary $a_{0}$. That is a physically expected result because $a_{0}$
can be always absorbed into the spacetime and does not provide any properties
in the solution. The values, $p_{1}$ and $p_{2}$, do not provide a singularity
for the field equations and so we select to proceed with the value $p_{3}$.
While the latter is a fraction we can work as it is and we do not have to do
any transformation $a\rightarrow a^{\lambda}$ in order to make the value of
$p$ negative.

The second step is to determine the resonances, $s$, which should be rational
numbers, i.e., $s\in%
%TCIMACRO{\U{211a} }%
%BeginExpansion
\mathbb{Q}
%EndExpansion
$. To do that we substitute%
\begin{equation}
a\left(  \tau\right)  =a_{0}\tau^{\frac{5}{6}}+\gamma\tau^{\frac{5}{6}%
+s}~,~R\left(  \tau\right)  =\frac{10}{3}\tau^{-2}+\delta\tau^{\frac{5}{6}+s}
\label{ac.07}%
\end{equation}
into (\ref{fr.07})-(\ref{fr.05}) where $\gamma,~\delta$ are two numbers which
we want to be arbitrary and so give the additional constants of integration of
the system. From this we find the following system%
\begin{equation}%
%TCIMACRO{\QATOPD{\{}{\}}{a_{0}\delta-2s\left(  7+3s\right)  \gamma=0}%
%{a_{0}\left(  25-8s+12s^{2}\right)  \delta-360s\gamma=0} }%
%BeginExpansion
\genfrac{\{}{\}}{0pt}{}{a_{0}\delta-2s\left(  7+3s\right)  \gamma
=0}{a_{0}\left(  25-8s+12s^{2}\right)  \delta-360s\gamma=0}
%EndExpansion
\label{ac.08}%
\end{equation}
which are required to have solution for the parameters, $\gamma,\delta$, to be
arbitrary. Hence the determinant of the system (\ref{ac.08}) has to be zero
and this gives the resonances\footnote{We note that expression (\ref{ac.07})
does not solve the field equations, that is, $R\left(  \tau\right)  $ in
(\ref{ac.07}) is not given from (\ref{fr.05}) for the scale factor $a\left(
\tau\right)  =a_{0}\tau^{\frac{5}{6}}+\gamma\tau^{\frac{5}{6}+s}$. The
analysis is equivalent either if we had applied it in the fourth-order
equation (\ref{fr.07}), where $R$ has been replaced from (\ref{fr.05}).
}%
\begin{equation}
s_{1}=-1~,~s_{2}=0~,~s_{3}=\frac{1}{6}~,~s_{4}=-\frac{5}{6}. \label{ac.09}%
\end{equation}

Hence, as all the resonances are rational numbers, we say that the model
(\ref{ac.05}) with $n=\frac{4}{3}$ passes the singularity test. Furthermore,
because one of the resonances is zero, the coefficient of the leading-order
term is arbitrary. As $s_{3},s_{4}$ are smaller numbers than $p_{3}$, the
Laurent expansion is a mixed\ (Left and Right) Painlev\'{e} Series. Hence the
solution of the scale factor is in the form
\begin{equation}
a\left(  \tau\right)  =%
%TCIMACRO{\dsum \limits_{J=-2}^{-\infty}}%
%BeginExpansion
{\displaystyle\sum\limits_{J=-2}^{-\infty}}
%EndExpansion
\alpha_{J}\tau^{\frac{5+J}{6}}+\alpha_{-1}\tau^{-\frac{4}{6}}+\alpha_{0}%
\tau^{\frac{5}{6}}+\alpha_{1}\tau^{\frac{4}{6}}+%
%TCIMACRO{\dsum \limits_{I=2}^{+\infty}}%
%BeginExpansion
{\displaystyle\sum\limits_{I=2}^{+\infty}}
%EndExpansion
\alpha_{I}\tau^{\frac{5+I}{6}} \label{ac.10}%
\end{equation}
in which the coefficient constants have to be determined by substitution into
the system (\ref{fr.07})-(\ref{fr.05}). There is also the last test that one
has to perform. This is the consistency of the solution when the nondominant
terms are considered. However, for this case we know well that the system in
integrable and it is not necessary to present the consistency test.
Furthermore from (\ref{ac.06}) and for the dominator $p_{4}$, we find the
following resonances $s_{1}=-1~,~s_{2}=0~,~s_{3}=-\frac{2}{9}~,~s_{4}%
=-\frac{7}{9},~$which give us the second solution~%
\begin{equation}
a\left(  \tau\right)  =\bar{\alpha}_{0}\tau^{\frac{8}{9}}+\bar{\alpha}_{1}%
\tau^{\frac{7}{9}}+\bar{\alpha}_{2}\tau^{\frac{6}{9}}+%
%TCIMACRO{\dsum \limits_{I=3}^{\infty}}%
%BeginExpansion
{\displaystyle\sum\limits_{I=3}^{\infty}}
%EndExpansion
\bar{\alpha}_{I}\tau^{\frac{8-I}{9}}, \label{ac.12}%
\end{equation}
which is a Left Painlev\'{e} Series in contrary to the other solution which is
expressed with a mixed Painlev\'{e} Series. The second solution is expected
and from the results of \cite{palcqg}. However, before we proceed with the
general case let us consider a second particular case of (\ref{ac.05}) in
which $n=\frac{3}{2}$.

For the latter value of $n$ the field equations do not admit any point
symmetry or contact symmetry. That does not mean that a higher-order symmetry
does not exist. In order to perform the singularity analysis we perform in
(\ref{fr.06})-(\ref{fr.05}) the change of variable $a\left(  t\right)
=\left(  b\left(  t\right)  \right)  ^{-1}$. Following the same steps we find
the dominant behaviour $\left(  b,R\right)  \rightarrow\left(  a_{0}\tau
^{p},R_{0}\tau^{q}\right)  $, for $R_{0}=6\left(  2p^{2}+p\right)  $ and
\begin{equation}
p_{1}=0~,~p_{2}=-\frac{1}{2}~,~p_{3}=-1~,~p_{4}=-2. \label{ac.12a}%
\end{equation}

We consider the dominant exponent, $p_{3}=-1$, for which the corresponding
resonances are $s_{1}=-1~,~s_{2}=0~,~s_{3}=-2~,~s_{4}=1$, that is, the system
passes the singularity test and it is integrable. The solution is given in a
form of the Laurent expansion,%
\begin{equation}
b\left(  \tau\right)  =%
%TCIMACRO{\dsum \limits_{J=-3}^{-\infty}}%
%BeginExpansion
{\displaystyle\sum\limits_{J=-3}^{-\infty}}
%EndExpansion
\beta_{J}\tau^{-1+J}+\beta_{-2}\tau^{-3}+\beta_{-1}\tau^{-2}+\beta_{0}%
\tau^{-1}+\beta_{1}+\beta_{2}\tau+%
%TCIMACRO{\dsum \limits_{I=3}^{+\infty}}%
%BeginExpansion
{\displaystyle\sum\limits_{I=3}^{+\infty}}
%EndExpansion
\beta_{I}\tau^{-1+I}, \label{ac.13}%
\end{equation}
and%
\begin{equation}
R\left(  \tau\right)  =r_{0}\tau^{-2}+r_{1}\tau^{-1}+r_{2}+%
%TCIMACRO{\dsum \limits_{I=3}^{\infty}}%
%BeginExpansion
{\displaystyle\sum\limits_{I=3}^{\infty}}
%EndExpansion
r_{I}\tau^{-2+I}. \label{ac.14}%
\end{equation}

In order to test the consistency of the solution (\ref{ac.13}), (\ref{ac.14}),
we substitute into (\ref{fr.06})-(\ref{fr.05}) where we derive the first
coefficients $\beta_{I},r_{I}$. Recall that, as one of the resonances is zero,
$\beta_{0}$ has to be arbitrary. We find the following values $\left(
r_{0},r_{1},r_{2},...\right)  =\left(  6,-24\frac{\beta_{1}}{\beta_{0}%
},6\left(  7\beta_{1}^{2}-10\beta_{0}\beta_{2}\right)  \beta_{0}%
^{-2},...\right)  $, and $\left(  \beta_{2},\beta_{3},\beta_{4},...\right)
=\frac{\beta_{1}}{\beta_{0}}\left(  \frac{3\beta_{1}}{4},\frac{\beta_{1}^{2}%
}{4\beta_{0}},-\frac{13}{32}\frac{\beta_{1}^{3}}{\beta_{0}^{2}},...\right)  $.
From this we observe that the constants $\beta_{0},~\beta_{1}~$are arbitrary,
the third integration constant is hidden in the other coefficients while the
fourth integration constant is the position of the singularity $\tau_{0}.$
Furthermore from the exponent, $p_{4}=-2$, we determine the second solution.

For the dominant behavior, $p_{4}=-2$, the corresponding resonances are
$s_{1}=-1$,~$s_{2}=0$,~$s_{3}=-3$~and $s_{4}=-4$. Hence the solution is%
\begin{equation}
b\left(  \tau\right)  =\bar{\beta}_{0}\tau^{-2}+\beta_{1}\tau^{-3}+\beta
_{2}\tau^{-4}+%
%TCIMACRO{\dsum \limits_{I=3}^{\infty}}%
%BeginExpansion
{\displaystyle\sum\limits_{I=3}^{\infty}}
%EndExpansion
\beta\tau^{-2-I} \label{ac.15}%
\end{equation}
and%
\begin{equation}
R\left(  \tau\right)  =\bar{r}_{0}\tau^{-2}+\bar{r}_{1}\tau^{-3}+r_{2}%
\tau^{-4}+%
%TCIMACRO{\dsum \limits_{I=3}^{\infty}}%
%BeginExpansion
{\displaystyle\sum\limits_{I=3}^{\infty}}
%EndExpansion
r_{I}\tau^{-2-I}, \label{ac.16}%
\end{equation}
which is a Left Painlev\'{e} Series. The first coefficients $\bar{\beta}_{I}$,
and $\bar{r}_{I}$ are: $\left(  \bar{r}_{0},\bar{r}_{1},\bar{r}_{2}%
,...\right)  =\left(  36,36\frac{\bar{\beta}_{1}}{\beta_{0}},-6\left(
3\bar{\beta}_{1}^{2}-10\bar{\beta}_{0}\bar{\beta}_{2}\right)  \bar{\beta}%
_{0}^{-2},...\right)  $ and $\left(  \bar{\beta}_{2},\bar{\beta}%
_{3},...\right)  =\frac{\bar{\beta}_{1}}{\beta_{0}}\left(  \frac{3\bar{\beta
}_{1}}{4},\frac{\bar{\beta}_{1}^{2}}{5\bar{\beta}_{0}},...\right)  $, which
pass the consistency test. Finally if we substitute the solution in
(\ref{fr.06}) then the relation among the integration constants and the matter
density $\rho_{m0}$ will be found.

We apply the same algorithm for general value $n$, and we find that there are
only two possible leading terms \ $p_{I}$ and $p_{II}$ which depend on the
value of $n.~$We summarize the results in the following proposition.

\textbf{Proposition:} \textit{The gravitational field equations,
\textit{(\ref{fr.06})-(\ref{fr.05}),} of the fourth-order theory of gravity
}$R+qR^{n}$\textit{, with }$q\neq0$\textit{, }$n\neq0,1$\textit{, pass the
singularity test for }$n>1~,~n\neq2,\frac{5}{4}$\textit{, in which the
following set of leading-order exponents and of the resonances are rational
numbers. }\newline\textit{I) For the dominant term }$p_{I}=\frac{2n^{2}%
-3n+1}{n-2}$\textit{ with }$p_{I}\in\left\{
%TCIMACRO{\U{211a} }%
%BeginExpansion
\mathbb{Q}
%EndExpansion
-%
%TCIMACRO{\U{2115} }%
%BeginExpansion
\mathbb{N}
%EndExpansion
\right\}  $\textit{ the corresponding resonances are }%
\begin{equation}
s_{1}=-1~,~s_{2}=0~,~s_{3}=\frac{5-14n+8n^{2}}{n-2}~,~s_{4}=1+s_{3}.
\label{ac.17}%
\end{equation}

\textit{while for }$p_{I}\in%
%TCIMACRO{\U{2115} }%
%BeginExpansion
\mathbb{N}
%EndExpansion
$\textit{ we make the change of variable }$a\left(  \tau\right)  \rightarrow
b^{-1}\left(  \tau\right)  $\textit{ in which the new dominant term is }%
$\bar{p}_{I}=-p_{I}$\textit{ and the resonances are the same.\qquad
\qquad\qquad\newline II) For the dominant term }$p_{II}$\textit{\thinspace
}$=\frac{2}{3}n$\textit{, in which }$p_{II}\in\left\{
%TCIMACRO{\U{211a} }%
%BeginExpansion
\mathbb{Q}
%EndExpansion
-%
%TCIMACRO{\U{2115} }%
%BeginExpansion
\mathbb{N}
%EndExpansion
\right\}  $\textit{ the corresponding resonances are}%
\begin{equation}
s_{1}=-1~,~s_{2}=0~,~s_{\pm}=\frac{3\left(  1-n\right)  \pm\sqrt{\left(
n-1\right)  \left(  256n^{3}-608n^{2}+417n-81\right)  }}{6\left(  n-1\right)
}. \label{ac.18}%
\end{equation}
\textit{while, when }$p_{II}\in%
%TCIMACRO{\U{2115} }%
%BeginExpansion
\mathbb{N}
%EndExpansion
$\textit{, we make the change of variable }$a\left(  \tau\right)  \rightarrow
b^{-1}\left(  \tau\right)  $\textit{, in which the new dominant term is }%
$\bar{p}_{II}=-p_{II}$\textit{ and the resonances are the same.}

It is important to mention that from (\ref{ac.18}), when $n$ is a rational
number, then the dynamical system is always integrable with leading term
$p_{I}$ The latter of course is not true for the exponents $p_{II}$

Furthermore the solution of the scale factor in the singularity is described
from the dominant term, that is, from the power-law solution $a\left(
t\right)  \simeq\tau^{p}$. which corresponds to a cosmological solution in
which \ the total fluid has a constant equation of state parameter
\begin{equation}
w_{tot}=\frac{p}{\rho}=-\frac{6n^{2}-11n+7}{3\left(  2n-1\right)  \left(
n-1\right)  }. \label{ac.19}%
\end{equation}

For the latter for the range of $n$ in which the field equations
(\ref{fr.06})-(\ref{fr.05}) pass the singularity analysis we have that
$w_{tot}<1$. Specifically $w_{tot}<-1,~$for $n\in\left(  1,2\right)  $ and
$w_{tot}>-1$ for $n>2$, while for large values of $n$, we have that
$\lim_{n\rightarrow\infty}w_{tot}$ $=-1^{+}$.

However, expression (\ref{ac.19}) is that of the $f\left(  R\right)  =$
$R^{n}$ gravity has been found \cite{palNEB}, which means that, when the
system (\ref{fr.06})-(\ref{fr.05}) reaches the singularity, then the model
(\ref{ac.05}) has the behaviour\footnote{The dynamics of that model can be
found in \cite{capRN}} of $R^{n}$.

The consistency test have been applied for two special values of $n$,~the
value $n_{1}=\frac{4}{3}$ and $n_{2}=\frac{3}{2}$, which both corresponds to
the two possible cases of\ the above proposition. To perform the consistency
test for general $n$ is difficult since the step on the Laurent expansion will
be unknown as also the position of the integration constants. These depend on
the value of $n$.

We introduced the singularity analysis of differential equations to study the
integrability of higher-order theories of gravity. Specifically we applied the
method on $f\left(  R\right)  $-gravity in the metric formalism. We considered
a special form of $f\left(  R\right)  $ which has been proposed in \cite{fr6}
and we found that for specific values of $n$, with $n>1$, the gravitational
field equations in a spatially FLRW spacetime are integrable. The singularity
analysis is complementary to the symmetry analysis. Symmetries, as we
discussed above, have been used for the determination of the unknown
potentials/models in different gravitational theories and consequently for the
determination of new integrable systems in cosmological studies. However, only
the point symmetries and the contact symmetries have been applied till now and
that does not mean that there are not systems which admit other kind of
symmetries, such as nonlocal symmetries or Lie-B\"{a}cklund symmetries which
in general are difficult to calculate. \

Furthermore for the case (I) of the above proposition we would like to give
the ranges of $n$, in which the analytical solution is expressed in a Right or
Left Painlev\'{e} Series. For values of $n$ in which $p_{I}$ is not a positive
integer number, for $n\in\left(  1,\frac{5}{4}\right)  \cup\left(  2,n\right)
$ the solution is expressed from a Right Painlev\'{e} Series and for
$n\in\left(  \frac{13}{6}+\sqrt{\frac{73}{6}},2\right)  ~$the solution is
expressed in a Left Painlev\'{e} Series, while for $n\in\left(  \frac{5}%
{4},\frac{13}{6}+\sqrt{\frac{73}{6}}\right)  ~$the Painlev\'{e} Series has
left and right expansions.

As we discussed above $f\left(  R\right)  $ gravity is equivalent with
O'Hanlon gravity, that is, the Action integral (\ref{ac.01}) can be written as
that of Brans-Dicke Action, with Brans-Dick parameter zero, $\omega_{BD}=0$,
where the latter is equivalent with a minimally coupled scalar field
cosmological model under a conformal transformation.

Consider that $\phi=f_{,R}$, then for (\ref{ac.05}) we find,~$R^{n-1}=\left(
\frac{1}{nq}\right)  \left(  \phi-1\right)  $, where in the case of vacuum
Lagrangian (\ref{ac.03}) becomes,
\begin{equation}
L\left(  N,a,\dot{a},R,\dot{R}\right)  =\frac{1}{N}\left(  6a\phi\dot{a}%
^{2}+6a^{2}\dot{a}\dot{\phi}\right)  +N\left(  a^{3}V_{0}\left(
\phi-1\right)  ^{\frac{n}{n-1}}\right)  ,\label{lan.01}%
\end{equation}
in which $V_{0}=q\left(  n-1\right)  \left(  nq\right)  ^{-\frac{1}{n-1}}$.
Consider $a\left(  t\right)  =A\left(  t\right)  \phi\left(  t\right)
^{-\frac{1}{2}}$, and $N\left(  t\right)  =\bar{N}\left(  t\right)
\phi\left(  t\right)  ^{-\frac{1}{2}}$, hence Lagrangian (\ref{lan.01})
becomes%
\begin{equation}
L\left(  \bar{N},A,\dot{A},\psi,\dot{\psi}\right)  =\frac{1}{\bar{N}}\left(
3A\dot{A}^{2}-\frac{1}{2}A^{3}\dot{\psi}^{2}\right)  +\bar{N}A^{3}\bar{V}%
_{0}\left(  e^{-\left(  \frac{4}{3\mu}+\frac{2}{3}\right)  \psi}-e^{-\frac
{4}{3\mu}\psi}\right)  ^{\mu}\label{lan.02}%
\end{equation}
where $\mu=\frac{n}{n-1}$, $V_{0}=2\bar{V}_{0}$~and the new field $\psi\left(
t\right)  $ is related with $\phi\left(  t\right)  $ with the formula $\ln
\phi=\frac{2}{3}\psi$. Therefore from (\ref{lan.02}) we have that the
potential of the minimally coupled scalar field is
\begin{equation}
V\left(  \psi\right)  =\bar{V}_{0}\left(  e^{-\left(  \frac{4}{3\mu}+\frac
{2}{3}\right)  \psi}-e^{-\frac{4}{3\mu}\psi}\right)  ^{\mu}\label{lan.03}%
\end{equation}
which is not included in the list of integrable potentials of \cite{fer}.

In a forthcoming work we wish to extend our analysis to other proposed models
which can be found in the literature and also to study the physical properties
of these models which are found to be integrable.

\begin{acknowledgments}
AP acknowledges financial support \ of FONDECYT grant no. 3160121. PGL Leach
thanks the Instituto de Ciencias F\'{\i}sicas y Matem\'{a}ticas of the UACh
for the hospitality provided while this work carried out and acknowledges the
National Research Foundation of South Africa and the University of
KwaZulu-Natal for financial support. The views expressed in this letter should
not be attributed to either institution.
\end{acknowledgments}

\end{document}